\documentstyle[preprint,prb,aps]{revtex}
\tighten
\begin{document}

\draft

\title{Multiplet Effects in the Quasiparticle Band Structure
of the $f^1-f^2$ Anderson Model}

\author{B.R. Trees}
\address{Department of Physics, Monmouth College,
Monmouth, IL 61462}
\author{A.J. Fedro}
\address{Physics Department, Northern Illinois University, DeKalb, IL  60115 \\
and Materials Science Division, Argonne National Laboratory,
Argonne, IL 60439}
\author{M.R. Norman}
\address{Materials Science Division, Argonne National Laboratory,
Argonne, IL 60439}

\date{\today}

\maketitle

\begin{abstract}

In this paper, we examine the mean field electronic structure of the
$f^1-f^2$ Anderson lattice model in a slave boson approximation, which should
be
useful in understanding the physics of correlated metals with more than one
f electron per site such as uranium-based heavy fermion superconductors.
We find that the multiplet structure of the $f^2$ ion
acts to quench the crystal field
splitting in the quasiparticle electronic structure.  This is consistent
with experimental observations in such metals as $UPt_3$.

\end{abstract}


\narrowtext

Localized electron systems exhibit multiplets of definite L, S, and J
due to the symmetry of the Coulomb interaction about an atomic site.  The
degeneracies of these multiplets are not properly described by standard
electronic structure techniques used in solids which assume the filling of
single particle levels using Fermi-Dirac statistics.  Despite this, a Landau
quasiparticle approach appears to be successful even in metals which exhibit
high energy multiplet excitations.  A good example of this is the heavy
fermion superconductor $UPt_3$.  This metal exhibits canonical Fermi liquid
behavior at low temperatures with a Fermi surface in reasonable agreement
with band calculations.\cite{dhva}  On the other hand, high energy
neutron scattering data reveal
an excitation at about 0.4 eV,\cite{osborn} close to the 0.5 eV transition
seen in the free ion, known
to be a multiplet excitation between the $^3H_4$ ground state and the $^3F_2$
excited state.  Therefore, $UPt_3$ is quite unusual in that its
low energy excitations exhibit single particle Kramers degeneracy whereas
its high energy ones exhibit multi-particle Coulomb degeneracy.  Our purpose
here is to understand what influence
these $f^2$ multiplet correlations have on the low energy quasiparticle
excitations.  We do this by studying the $f^1-f^2$ Anderson lattice model at
the mean field level within a slave boson approximation.  Since
this study deals with the relation of single particle statistics to
multi-particle statistics, it should be of relevance to other problems as well.

The $f^1-f^2$ Anderson model has been investigated by a number of authors.
Early work studied the single impurity model
either from a variational approach\cite{yvj,nrg} or by use of the 1/N
expansion\cite{rdrn} where N is the degeneracy of the ion.  The lattice case
was in turn looked at by various authors within a slave boson
approximation,\cite{rh,eg} but the effects of multiplets were ignored.
Multiplet notation was introduced in what we will denote the $mn$ basis by
Newns and Read,\cite{adv,newns} but no calculations were carried out explicitly
including multiplet effects (the $mn$ state has one f electron in
level
m and the other in level n).
This work was extended by Dorin and Schlottmann\cite{ds}
who dealt with the general $f^n-f^{n+1}$ problem in the $mn$ basis and also
included Gutzwiller corrections to the mean field approximation, which are
necessary when looking for possible magnetic instabilities.\cite{krab}  Our
treatment in the $mn$ basis will closely parallel this work.  In turn, though,
we will develop the formalism in the $JJ_z$ basis, where J is the total
angular momentum of the $f^2$ state.  This basis is diagonal in the Coulomb
interaction and contains the proper multiplet degeneracy of the $f^2$ ion.

The ionic part of the Hamiltonian in the $mn$ basis is
\begin{equation}
\sum_{jm} E_m|m><m| + \sum_{jmn} E_{mn}|mn><mn|
\label{eq:1}
\end{equation}
where $E_m$ is the energy of one f electron in level m, $E_{mn}$
is the energy of two f electrons in levels m and n, and j is the site index.
Note that the state $|mn>$ is antisymmetric in
the two indices.  We will now assume that we are in the strong spin-orbit
coupling limit (as appropriate for uranium) so that the $f^1$ state has
a J of 5/2 (6 states) and the $f^2$ state has a J of 0, 2, or 4
(15 states).

The conduction electron part of the Hamiltonian is
\begin{equation}
\sum_{\vec k m} \epsilon_{\vec k m} c^{\dagger}_{\vec k m}
c_{\vec k m}
\label{eq:2}
\end{equation}
where m is assumed to run over the same 6 indices as the $f^1$ state.

The hybridization part of the Hamiltonian is
\begin{equation}
\sum_{j\vec k mn} V_{\vec k m} e^{i\vec k \cdot \vec R_j}
|mn><n| c_{\vec k m} + H.C.
\label{eq:3}
\end{equation}
where H.C. denotes the Hermitian conjugate.  In this paper we replace
$V_{\vec k m}$ by a constant, $V$.

To solve this Hamiltonian, we employ the slave boson approximation.  We
associate a boson $p_{jm}$ at each site j for each $f^1$ state in level m
and a boson $d_{jmn}$ at each site j for each $f^2$ state in level mn.
For the mean field approximation we consider, these bosons are replaced by c
numbers assumed to be site independent.
The ionic part of the Hamiltonian per site can now be written as
\begin{equation}
\sum_{m} E_m p^2_m + \sum_{mn} E_{mn} d^2_{mn}
\label{eq:4}
\end{equation}
We must also satisfy a completeness relation
\begin{equation}
\sum_m p^2_m + \sum_{mn} d^2_{mn} = 1
\label{eq:5}
\end{equation}
We associate a Lagrange multiplier $\lambda^1$ with this constraint.  The other
constraints involve the f count in each channel $m$ at each site
\begin{equation}
h_m \equiv f^{\dagger}_{m} f_{m} = p^2_m + \sum_n d^2_{mn}
\label{eq:6}
\end{equation}
with which we associate Lagrange multipliers $\lambda_m$.

Finally, the hybridization term now becomes
\begin{equation}
\sum_{\vec k m} V f^{\dagger}_{\vec k m} c_{\vec k m} z_m + H.C.
\label{eq:7}
\end{equation}
where
\begin{equation}
z_m = \sum_n d_{mn} p_n
\label{eq:8}
\end{equation}
and $f_{\vec k m}$ is the Fourier transform of $f_{jm}$.  The Gutzwiller
generalization would involve replacing $z_m$ by\cite{ds}
\begin{equation}
\tilde z_m = \frac{z_m}{\sqrt{(1-h_m)h_m}}
\label{eq:9}
\end{equation}
We do not use such a renormalization here, but it can be easily included.

We now have to determine the following parameters: 6 $p_m$, 15 $d_{mn}$,
$\lambda^1$, and 6 $\lambda_m$.  These are determined by
setting the derivative of the Hamiltonian with respect to each of them
equal to zero.  The final
parameter is the chemical potential, $\mu$, determined by
the total number of electrons.

The quasiparticle bands which come out of this model are the standard ones
\begin{equation}
E^{\pm}_{\vec k m} = \frac{\lambda_m + \epsilon_{\vec k m}}{2}
\pm \frac{1}{2}\sqrt{(\lambda_m - \epsilon_{\vec k m})^2 + 4 z_m^2 V^2}
\label{eq:10}
\end{equation}
The following mean field relations are necessary to
solve these equations (we assume a constant conduction electron density
of states, $N_0$)\cite{adv}
\begin{equation}
\sum_{\vec k}<c^{\dagger}_{\vec k m}c_{\vec k m}>
 = N_0 (\mu - \epsilon_a^m)
\end{equation}
\begin{equation}
\sum_{\vec k}<f^{\dagger}_{\vec k m}f_{\vec k m}> = N_0 z_m^2 V^2
 (\frac{1} {\epsilon_a^m - \lambda_m} - \frac{1}{\mu - \lambda_m})
\end{equation}
\begin{equation}
\sum_{\vec k}Re<f^{\dagger}_{\vec k m}c_{\vec k m}> = N_0 z_m V
 \ln(\frac{\mu - \lambda_m}{\epsilon_a^m - \lambda_m}) \\
\end{equation}
which we denote by $I_m$, $J_m$, and $K_m$ respectively
where $\epsilon_a^m$ is the bottom of the $E^-_m$ band.  These equations
as written assume only the $E^-_m$ bands are occupied for notational
simplicity, but in actual calculations, we used the full expressions
which can include contributions from the $E^+_m$ bands depending on the
value of the chemical potential, $\mu$.  We note that the f count,
$n_f$, is simply the sum over $m$ of the $J_m$.

The $mn$ basis is not diagonal in the Coulomb interaction, so
to treat the effect of multiplets properly, we convert the above
formalism to the $JJ_z$ basis.  We note that
\begin{equation}
|JJ_z> = \sum_{mn} U^{mn}_{JJ_z} |mn>
\label{eq:11}
\end{equation}
where the $U$ are Clebsch-Gordon coefficients.\cite{co}
These act to define a new boson field for the $f^2$ state which we denote
as $d_{JJ_z}$.

The ionic part of the Hamiltonian per site is now
\begin{equation}
H_{ionic} = \sum_{m} E_m p^2_m + \sum_{JJ_z} E_{JJ_z} d^2_{JJ_z}
\label{eq:12}
\end{equation}
with the completeness relation
\begin{equation}
\sum_m p^2_m + \sum_{JJ_z} d^2_{JJ_z} = 1
\label{eq:13}
\end{equation}
To convert Eqs. 6 and 8, though, one must remember that
$d$ is actually an operator and that $d_{mn}$
and $d_{JJ_z}$ represents its expectation value in each basis.  Thus,
taking advantage of the assumed diagonality in the $JJ_z$ basis, we have
\begin{equation}
d_{mn} = \sum_{JJ_z}U^{mn}_{JJ_z} d_{JJ_z} U^{mn}_{JJ_z}
\end{equation}
\begin{equation}
d^2_{mn} = \sum_{JJ_z} U^{mn}_{JJ_z} d^2_{JJ_z} U^{mn}_{JJ_z}
\end{equation}
These two equations are then inserted into Eqs. 6 and 8
to obtain the values of $h_m$ and $z_m$ in the $JJ_z$ basis.

Solving the full problem above is quite difficult given the large
number of variational parameters.  We will therefore consider
a simple model which illustrates the relevant points.  In this model,
we only keep one $f^2$ configuration, the doublet $J,J_z = 4,\pm 4$.
Since the $f^1$ state $|\pm 1/2>$ does not couple to this, we elect to keep
only the $|\pm 3/2>$ and $|\pm 5/2>$ states in the $f^1$ and conduction
electron manifolds, thus further reducing the number of variational
parameters.  Since we are considering an unpolarized solution,
one has only the following equations to solve
\begin{eqnarray}
2(p_{3/2}^2+p_{5/2}^2 + d_{44}^2) - 1 = 0 \\
p_{5/2}^2+d_{44}^2 - J_{5/2} = 0 \\
p_{3/2}^2+d_{44}^2 - J_{3/2} = 0 \\
(E_{44}+\lambda^1-\lambda_{5/2}-\lambda_{3/2})d_{44} \nonumber \\
+ V(K_{5/2}p_{3/2} + K_{3/2}p_{5/2}) = 0 \\
(E_{5/2}+\lambda^1-\lambda_{5/2})p_{5/2} + VK_{3/2}d_{44} = 0 \\
(E_{3/2}+\lambda^1-\lambda_{3/2})p_{3/2} + VK_{5/2}d_{44} = 0 \\
2(J_{3/2}+J_{5/2}+I_{3/2}+I_{5/2}) - N_t = 0
\end{eqnarray}
with $z_{5/2} = d_{44}p_{3/2}$, $z_{3/2} = d_{44}p_{5/2}$,
and $N_t$ the total number of electrons.
We solve these seven equations by squaring them all and adding them together.
The resulting master equation is a function of seven variables and is minimized
using Powell's method.\cite{nrec}

For the results presented, we took a conduction band width of 2 eV, a
hybridization potential, $V$, of 0.7 eV, an $E_{5/2}$ of -0.5 eV, an
$E_{3/2}$ of -0.4 eV, and a total electron count of 3.5.  $E_{44}$ was varied
in steps of 0.05 eV.  Stable solutions were found for $E_{44}$ 0.13 eV and
lower.  For higher values of $E_{44}$, the minimiziation failed to converge
from a typical input.
A solution at 0.15 eV was generated by trying a large number of input
parameters.  For 0.20 eV, the lowest minimum was found for an $n_f$
of 1.  For low values of $E_{44}$, one asymptotically approaches an $n_f$
of 2, so the runs were terminated at -1.50 eV.

The results are presented in the three figures.  Fig. 1 shows a plot of the
renormalized crystal field splitting, $\Delta \lambda \equiv \lambda_{3/2}
-\lambda_{5/2}$, versus
$n_f$.  The data for $n_f > 1.2$ appears to follow a straight line which goes
to zero at $n_f = 2$ and interpolates to half the bare crystal field splitting
at $n_f = 1$.  Below an $n_f$ of 1.2, the chemical potential jumps from the
lower to the upper 5/2 band.  After this, the crystal field splitting begins
to approach the unrenormalized value of 0.1 eV (the lowest
minimum for $E_{44}$ of 0.20 eV had an $n_f$ of 1 and an unrenormalized
crystal field splitting).  We contrast this behavior
with the $f^0-f^1$ Anderson lattice model, where no renormalization of the
crystal field splitting occurs.  In Fig. 2, a plot of $\lambda_{5/2}-\mu$
is shown, which should be an estimate of the Kondo temperature for
this model (the absolute value is shown, since this quantity changes sign near
an $n_f$ of 1.2).  This vanishes linearly with $n_f$
as $n_f$ approaches 2 but
vanishes faster than this when $n_f$ approaches 1, with the latter behavior
beginning when $n_f$ drops below 1.2.  This same type of
asymmetry was seen in earlier impurity calculations.\cite{rdrn}  In Fig. 3,
we show the f occupation in each channel as a function of $n_f$.  Again,
a change in behavior occurs when $n_f$ drops below 1.2 with very
different limiting behavior as $n_f$ approaches 1 or 2.  We suggest, then,
that the jump of the chemical potential is necessary for
the solution to cross over from the behavior characteristic of $n_f$ near 2
to obtain the unrenormalized behavior as $n_f$ approaches 1.  The value of
$n_f$ at which this occurs is model dependent.

We now present a derivation of the linear behavior of Fig. 1.
Subtracting Eq. 24 from Eq. 23
and utilizing the defintion of $K_m$ in Eq. 13, we have
\begin{equation}
\Delta \lambda = \Delta \lambda_0 + N_0V^2d_{44}^2
\ln[\frac{\lambda_{3/2}-\mu}
{\lambda_{5/2}-\mu}\frac{\epsilon_a^{5/2}-\lambda_{5/2}}{\epsilon_a^{3/2}
-\lambda_{3/2}}]
\end{equation}
where $\Delta \lambda = \lambda_{5/2}-\lambda_{3/2}$ and
$\Delta \lambda_0 = E_{5/2}-E_{3/2}$.  As $n_f$ approaches 2, we can set
the second fraction in the argument of the logarithm to 1, and expanding the
logarithm, we obtain
\begin{equation}
\Delta \lambda \simeq \Delta \lambda_0 - N_0V^2d_{44}^2
\frac{\Delta \lambda}{\lambda_{av}-\mu}
\end{equation}
where $\lambda_{av} = \frac{1}{2}(\lambda_{5/2}+\lambda_{3/2})$.  This
approximation has been checked numerically and works well for $n_f$ near 2.
To get an expression for $\lambda_{av}-\mu$, we analyze the definition of
$J_m$ in Eq. 12.  Since $z_m$ is vanishing as $n_f$
approaches 2, this forces $\lambda_m$ to approach $\mu$, and so we can
ignore the first term in parenthesis in Eq. 12.  Using Eqs. 20 and 21, we
obtain
\begin{equation}
\lambda_{av}-\mu \simeq N_0V^2d_{44}^2\frac{p_{av}^2}{p_{av}^2+d_{44}^2}
\end{equation}
where $p_{av}^2 = \frac{1}{2}(p_{5/2}^2+p_{3/2}^2)$.  From Eqs. 19-21,
$4 p_{av}^2 = 2-n_f$, so
\begin{equation}
\lambda_{av}-\mu \simeq N_0V^2d_{44}^2\frac{2-n_f}{n_f}
\end{equation}
Substituting this into Eq. 27, we obtain
\begin{equation}
\Delta \lambda \simeq \frac{\Delta \lambda_0}{2}(2-n_f)
\end{equation}
which is the desired relationship.

Fig. 1 is the main result of this paper.  We note that except for the
special cases of the $J=4, J_z = \pm 4, \pm 3$ states, all $f^2$
states involve combinations of states in all three of the $f^1$ subbands.
Because of this, in general, all crystal field splittings in the
quasiparticle electronic structure should be
expected to vanish in the limit as $n_f$ approaches 2 in the paramagnetic
case.  In fact, this effect
should occur in the Kondo limit for any $f^n-f^{n+1}$ problem with
$n \neq 0$ (the $n=0$ case is special since the $f^0$ state has zero f
electrons).  We suggest this
as a possible reason for the lack of any definite evidence for crystal
field splittings in $UPt_3$.  Evidence for crystal field effects have been
seen in other heavy fermion superconductors ($URu_2Si_2, UPd_2Al_3$), but
these are a reflection of the crystal field splitting of the $f^2$ manifold,
not of the quasiparticles states.  We note that in these cases, the ground
state is magnetic, causing a profound
effect on the quasiparticle electronic structure, so the above considerations
based on a paramagnetic solution
would not apply (a weak moment state is found in $UPt_3$, but it does not
appear to affect the quasiparticle electronic structure).

In conclusion, by studying the $f^1-f^2$ Anderson lattice model, we find that
multi-particle correlations have a profound impact on the quasiparticle
electronic structure.  The results of this paper should have particular bearing
on uranium heavy fermion metals where such multi-particle ground states are
known to exist.

\acknowledgments

This work was supported by the U.S. Dept. of Energy,
Basic Energy Sciences, under Contract No. W-31-109-ENG-38.

\vfill\eject

\begin{figure}
\caption{Renormalized crystal field splitting, $\Delta \lambda$, versus $n_f$,
with a bare splitting of 0.1 eV.
The straight line is an analytic approximation valid near an $n_f$ of 2
derived in the text.  The change in behavior near an $n_f$ of 1.2 is due to
a jump in the chemical potential from the lower to the upper 5/2 quasiparticle
band.}
\label{fig1}
\end{figure}

\begin{figure}
\caption{Absolute value of the separation of the renormalized f level position
in the 5/2 channel from the chemical potential versus $n_f$, which should be
an estimate of the Kondo temperature.}
\label{fig2}
\end{figure}

\begin{figure}
\caption{f electron counts in the 3/2 and 5/2 channels versus $n_f$.}
\label{fig3}
\end{figure}

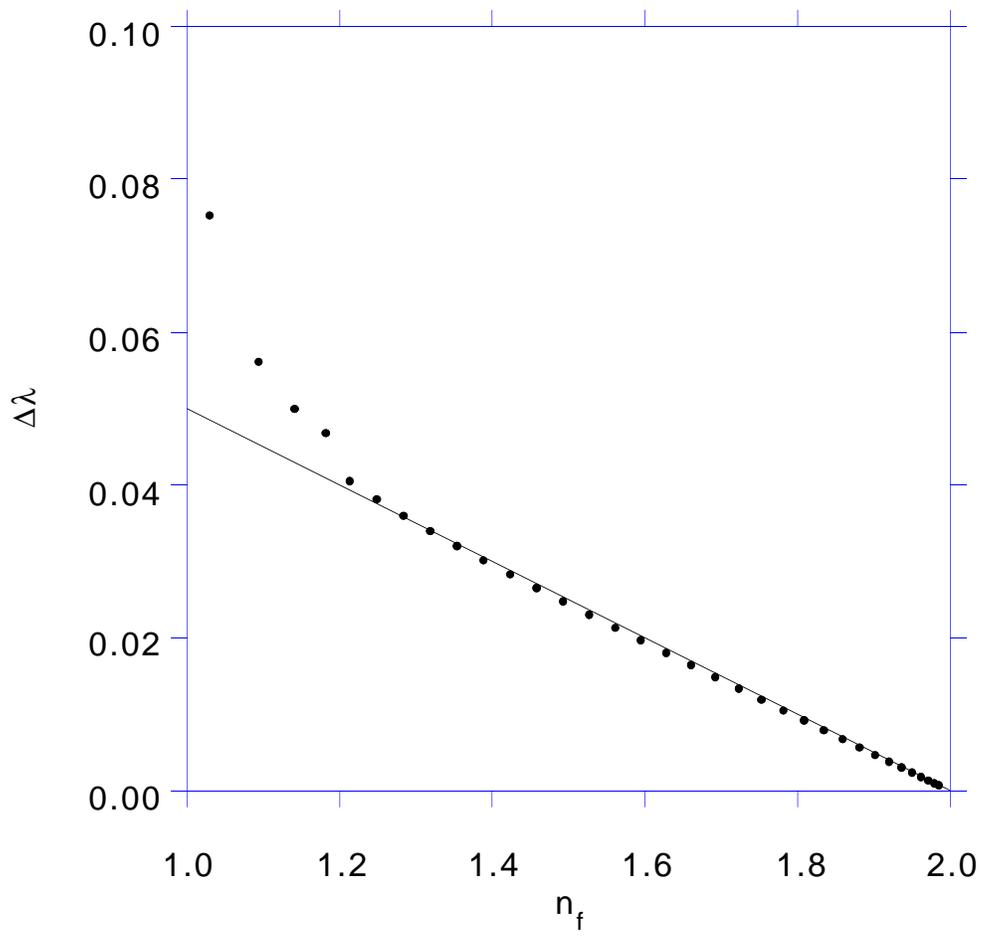
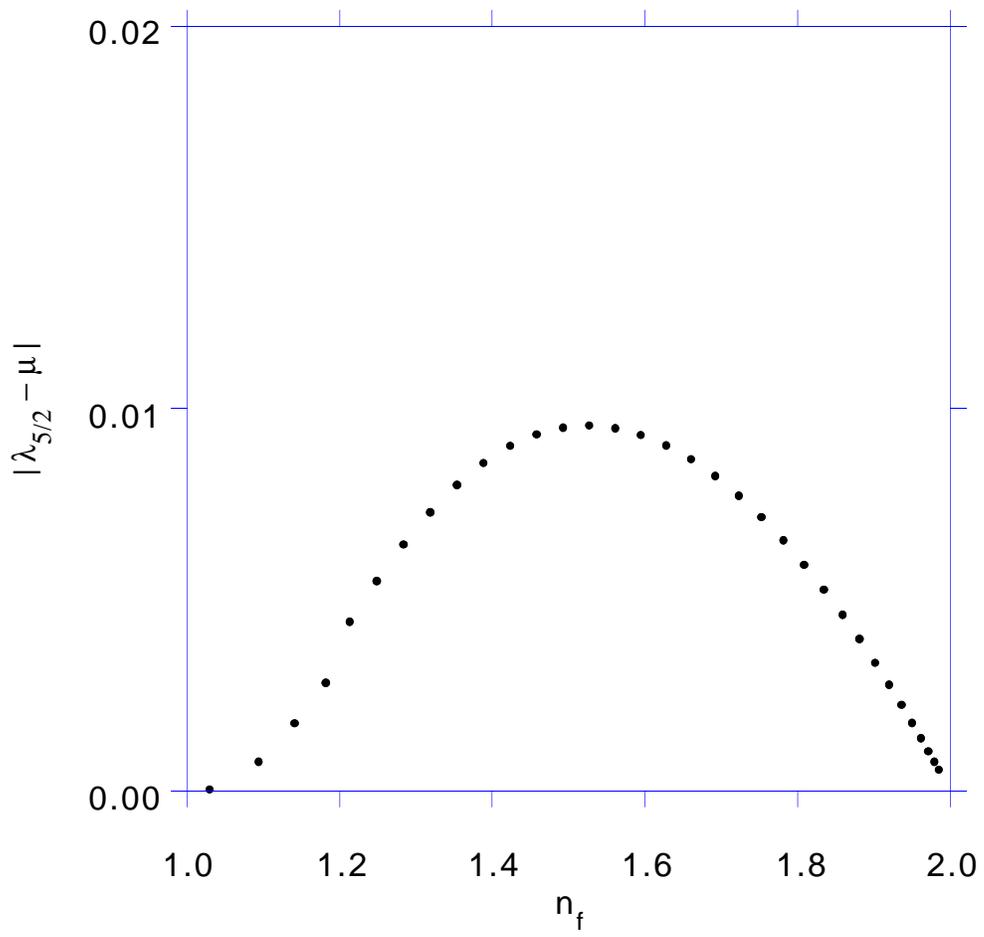
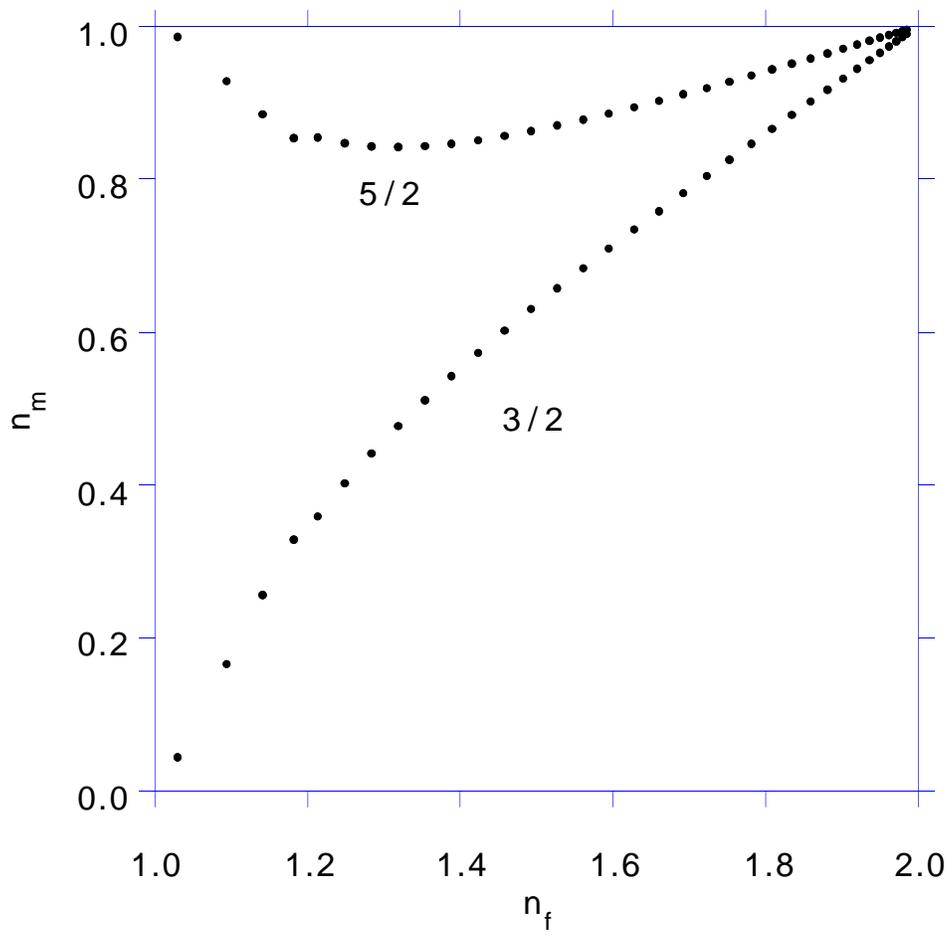
\end{document}